\documentclass{newaa} 
\usepackage{psfig,natbib}

\bibpunct{(}{)}{;}{a}{}{,} 

\begin{document}

\title{The CH out$-$of$-$plane bending modes of PAH molecules in
  astrophysical environments
  \thanks{based on observations
    obtained with ISO, an ESA project with instruments
    funded by ESA Member states (especially the PI
    countries: France, Germany, the Netherlands and the
    United Kingdom) and with the participation of ISAS and
    NASA.}}
\author{
  S. Hony\inst{1},
  C. Van Kerckhoven \inst{2},
  E. Peeters \inst{3,4},
  A.G.G.M Tielens \inst{3,4},
  D.M. Hudgins \inst{5},
  L.J. Allamandola \inst{5}
  }
\authorrunning{Hony et al.}
\titlerunning{CH OOP bending modes}
\offprints{S. Hony (hony@astro.uva.nl)} 
\institute{ Astronomical Institute `Anton Pannekoek',
  Kruislaan 403, 1098 SJ Amsterdam, The Netherlands \and
  Instituut voor Sterrenkunde, K.U. Leuven,
  Celestijnenlaan 200B, 3001 Heverlee, Belgium \and
  SRON Laboratory for Space Research Groningen, 
  P.O. Box 800, 9700 AV Groningen, The Netherlands \and
  Kapteyn Astronomical Institute PO Box 800, 9700 AV
  Groningen, The Netherlands \and
  NASA/Ames Research Center, MS:245-6, Moffett Field, CA
  94035-1000, U.S.A.}
\date{Received July 31, 2000; accepted July 31, 2000}
\abstract{ We present 10$-$15 $\mu$m spectra of a sample of H~{\sc ii}
  regions, YSOs and evolved stars that show strong unidentified
  infrared emission features, obtained with the ISO/SWS spectrograph
  on-board ISO. These spectra reveal a plethora of emission features
  with bands at 11.0, 11.2, 12.0, 12.7, 13.5 and 14.2 $\mu$m. These
  features are observed to vary considerably in relative strength to
  each-other from source to source. In particular, the 10$-$15 $\mu$m
  spectra of the evolved stars are dominated by the 11.2 $\mu$m band
  while for H~{\sc ii} regions the 12.7 is typically as strong as the
  11.2 $\mu$m band. Analysing the ISO data we find a good correlation
  between the 11.2 $\mu$m band and the 3.3 $\mu$m band, and between
  the 12.7 $\mu$m and the 6.2 $\mu$m band. There is also a correlation
  between the ratio of the UIR bands to the total dust emission and
  the 12.7 over 11.2 $\mu$m ratio.  Bands in the 10$-$15 $\mu$m
  spectra region are due to CH out$-$of$-$plane (OOP) bending modes of
  polycyclic aromatic hydrocarbons (PAHs). We summarise existing
  laboratory data and theoretical quantum chemical calculations of
  these modes for neutral and cationic PAHs.  Due to mode coupling,
  the exact peak position of these bands depends on the number of
  adjacent CH groups and hence the observed interstellar 10$-$15
  $\mu$m spectra can be used to determine the molecular structure of
  the interstellar PAHs emitting in the different regions. We conclude
  that evolved stars predominantly inject compact ~100$-$200 C-atom
  PAHs into the ISM where they are subsequently processed, resulting
  in more open and uneven PAH structures.  
  }
\maketitle

\section{Introduction}
Many sources show a rich set of emission features at 3.3, 6.2, 7.7,
8.6 and 11.2 $\mu$m commonly called the unidentified infrared (UIR)
emission features. These bands dominate the IR spectra of a wide
variety of objects, including H~{\sc ii} regions, PNe, post-AGB
objects, YSOs, the diffuse ISM, galaxies and galactic nuclei
\citep{1999ESASP.427.....C}. Understanding these features and the
origin and evolution of their carriers has become an important problem
in astrophysics. These UIR bands are generally attributed to IR
fluorescence of small ($\sim$ 50 C-atom) polycyclic aromatic
hydrocarbons (PAHs) molecules.

Laboratory spectroscopy of PAHs shows that, besides the well known UIR
bands, PAHs exhibit many weaker emission bands. In particular, the
region from 10 to 15 $\mu$m has a rich spectrum due to the
out$-$of$-$plane bending vibrations (OOP) of aromatic H-atoms. The
peak wavelength of these modes depends on the structure of the
molecule; in particular on the number of neighbouring H-atoms per ring
\citep[e.g.][]{1958Bellamy,1999ApJ...516L..41H}. Here we present data
in this region obtained with the Short Wavelength Spectrometer (SWS)
\citep{1996A&A...315L..49D} on-board the Infrared Space Observatory
(ISO) \citep{1996A&A...315L..27K}. The sensitivity and medium
resolving power of the instrument allows us to detect and resolve
several weak features predicted by the PAH hypothesis and to determine
the molecular structure of the emitting PAHs and their evolution in
space.

In Sect.~\ref{SecObservations}, we present the observations of our
sample of H~{\sc ii} regions, YSOs, reflection nebulae (RNe) and
evolved objects. The 10$-$15 $\mu$m regions of these sources are
analysed in Sect.~\ref{SecResults}. The spectral characteristics of
PAHs in this wavelength range as measured in the laboratory and
calculated by quantum chemical theories are summarised in
Sect.~\ref{SecLabdata}.  In Sect.~\ref{SecCompareLabISM} the
laboratory spectra are compared to the observed spectra. The molecular
structures implied by the observed spectra are discussed in
Sect.~\ref{SecMolstructure} while in Sect.~\ref{SecDiscussion} the
origin and evolution of these molecular structures are examined.
Finally in Sect.~\ref{SecSummary} our main results are summarised.

\section{Observations}
\label{SecObservations}
\begin{figure}
  \psfig{figure=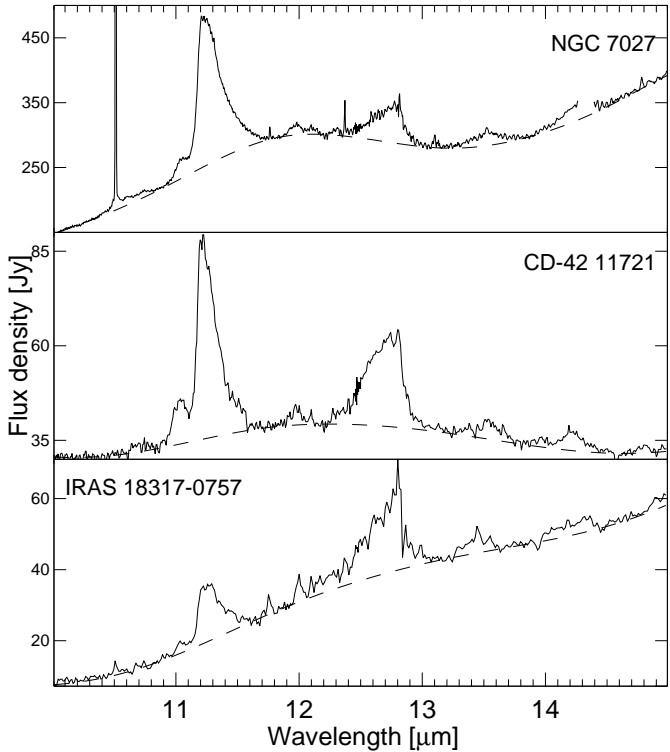,angle=0}
  \caption{Spectra of 3 sources that show features in the region
    of interest. The dashed lines are the continua mentioned in
    the text.}
  \label{FigFull}
\end{figure}
\begin{table*}
  \caption{
    Source list. Observational details of the sources used in this study.}
  \label{tab:obslog}
  \begin{tabular}{l l r r r r r l}
    \hline
    \hline
\multicolumn{1}{c}{$Object$}& 
\multicolumn{1}{c}{$Obs.^{a}$}& 
\multicolumn{1}{c}{$\alpha$}&
\multicolumn{1}{c}{$\delta$}& 
\multicolumn{1}{c}{$TDT^{b}$}&
\multicolumn{1}{c}{$Sp.Type$}&
\multicolumn{1}{c}{$G_\mathrm{0}$}&
\multicolumn{1}{c}{$Obj.Type$} 
\\
\multicolumn{1}{c}{$$}&
\multicolumn{1}{c}{$Mode$}&
\multicolumn{1}{c}{$(J2000)$}&
\multicolumn{1}{c}{$(J2000)$}&
\multicolumn{1}{c}{$$}& 
\multicolumn{1}{c}{$(K)$}& 
\multicolumn{2}{l}{$(1.6\,10^{-6}W/m^2)$}
\\
\hline
\object{AFGL 437}        &  01(2) &  03\,07\,23.68 &  $+$58\,30\,50.62 & 86300810  & O8.5    & 1E5     & Star forming region\\
\object{IRAS 03260+3111} &  01(3) &  03\,29\,10.37 &  $+$31\,21\,58.28 & 65902719  & B9      & 2E4     & Herbig AeBe\\ 
\object{NGC 2023}        &  01(3) &  05\,41\,38.30 &  $-$02\,16\,32.59 & 65602309  & B1.5V   & 3E2     & Refl. Nebula\\
\object{HD 44179}        &  01(4) &  06\,19\,58.20 &  $-$10\,38\,15.22 & 70201801  & B8V     & 5E6     & post-AGB\\
\object{IRAS 07027-7934} &  01(2) &  06\,59\,26.30 &  $-$79\,38\,48.01 & 73501035  & WC10    & 2E7     & PN\\
\object{HD 97048}        &  01(4) &  11\,08\,04.61 &  $-$77\,39\,16.88 & 61801318  & A0      & 2E4     & Herbig AeBe\\
\object{IRAS 12405-6238} &  01(3) &  12\,43\,31.93 &  $-$62\,55\,11.39 & 29400410  & O9.5    & 3E5     & H~{\sc ii}\\
\object{HEN 2-113}$^{\dagger}$   &&                &                   &           & WC10    & 6E4     & PN\\
        $-$              &  01(1) &  14\,59\,53.49 &  $-$54\,18\,07.70 & 07903307  &         &         & \\
        $-$              &  01(2) &  14\,59\,53.49 &  $-$54\,18\,07.70 & 43400768  &         &         & \\
\object{IRAS 15384-5348} &  01(2) &  15\,42\,17.16 &  $-$53\,58\,31.51 & 29900661  &         & 1E4     & H~{\sc ii}\\
        CD-42 11721(off) &  01(2) &  16\,59\,05.82 &  $-$42\,42\,14.80 & 28900461  & B0      & $-$     & Herbig AeBe (off pointing)\\
\object{CD-42 11721}$^{\dagger}$ &&                &                   &           & B0      & $-$     & Herbig AeBe\\
        $-$              &  01(2) &  16\,59\,06.82 &  $-$42\,42\,07.60 & 08402527  &         &         & \\
        $-$              &  01(2) &  16\,59\,06.80 &  $-$42\,42\,07.99 & 64701904  &         &         & \\
\object{IRAS 17047-5650}$^{\dagger}$&&             &                   &           & WC10    & 5E6     & PN\\
        $-$              &  01(3) &  17\,09\,00.91 &  $-$56\,54\,47.20 & 13602083  &         &         & \\
        $-$              &  01(1) &  17\,09\,00.91 &  $-$56\,54\,48.10 & 27301339  &         &         & \\
\object{HB 5}            &  01(3) &  17\,47\,56.11 &  $-$29\,59\,39.70 & 49400104  &120\,000$\ddagger$& $-$     & PN\\
\object{NGC 6537}        &  01(3) &  18\,05\,13.14 &  $-$19\,50\,34.51 & 70300475  & A0      & $-$     & PN\\
\object{GGD 27-ILL}$^{\dagger}$  &&                &                   &           & B1      & 3E6     & Star forming region\\
        $-$              &  01(2) &  18\,19\,12.04 &  $-$20\,47\,30.98 & 14802136  &         &         & \\
        $-$              &  01(2) &  18\,19\,12.03 &  $-$20\,47\,30.59 & 14900323  &         &         & \\
\object{IRAS 18240-0244} &  01(1) &  18\,26\,40.00 &  $-$02\,42\,56.99 & 14900804  & WC8     & $-$     & PN\\
\object{IRAS 18317-0757} &  01(2) &  18\,34\,24.94 &  $-$07\,54\,47.92 & 47801040  & O8      & $-$     & H~{\sc ii}\\
\object{IRAS 18416-0420} &  01(2) &  18\,44\,15.19 &  $-$04\,17\,56.40 & 13402168  & O5.5    & $-$     & H~{\sc ii}\\
\object{IRAS 18502+0051} &  01(2) &  18\,52\,50.21 &  $+$00\,55\,27.59 & 15201645  & O7      & $-$     & H~{\sc ii}\\
\object{TY CRA}$^{\dagger}$      &&                &                   &           & B9      & 6E3     & Herbig AeBe\\
        $-$              &  01(3) &  19\,01\,40.71 &  $-$36\,52\,32.48 & 33400603  &         &         & \\
        $-$              &  01(1) &  19\,01\,40.70 &  $-$36\,52\,32.59 & 34801419  &         &         & \\
        $-$              &  01(3) &  19\,01\,40.71 &  $-$36\,52\,32.48 & 71502003  &         &         & \\
\object{BD +30 3639}     &  01(3) &  19\,34\,45.20 &  $+$30\,30\,58.79 & 86500540  & WC9     & 1E5     & PN\\
\object{IRAS 19442+2427} &  01(2) &  19\,46\,20.09 &  $+$24\,35\,29.40 & 15000444  & O7      & 6E6     & H~{\sc ii}\\
\object{BD+40 4124}      &  01(3) &  20\,20\,28.31 &  $+$41\,21\,51.41 & 35500693  & B2      & 1E4     & Herbig AeBe\\
\object{S106 IRS4}       &  01(2) &  20\,27\,26.68 &  $+$37\,22\,47.89 & 33504295  & 08      & 1E5     & H~{\sc ii}\\
\object{NGC 7023}$^{\dagger}$    &&                &                   &           & B3      & 5E2     & Refl. Nebula\\
        $-$              &  01(4) &  21\,01\,31.90 &  $+$68\,10\,22.12 & 20700801  &         &         & \\
        $-$              &  01(2) &  21\,01\,30.40 &  $+$68\,10\,22.12 & 48101804  &         &         & \\
\object{NGC 7027}$^{\dagger}$    &&                &                   &           &200\,000$\diamond$& 2E5     & PN\\
        $-$              &  01(4) &  21\,07\,01.71 &  $+$42\,14\,09.10 & 02401183  &         &         & \\
        $-$              &  01(1) &  21\,07\,01.70 &  $+$42\,14\,09.10 & 23001356  &         &         & \\
        $-$              &  01(2) &  21\,07\,01.70 &  $+$42\,14\,09.10 & 23001357  &         &         & \\
        $-$              &  01(3) &  21\,07\,01.70 &  $+$42\,14\,09.10 & 23001358  &         &         & \\
        $-$              &  06    &  21\,07\,01.50 &  $+$42\,14\,10.00 & 33800505  &         &         & \\
        $-$              &  01(4) &  21\,07\,01.63 &  $+$42\,14\,10.28 & 55800537  &         &         & \\
\object{IRAS 21190+5140} &  01(2) &  21\,20\,44.85 &  $+$51\,53\,26.59 & 15901853  & $-$     & 2E5     & H~{\sc ii}\\
\object{IRAS 21282+5050}$^{\dagger}$&&             &                   &           & O9      & 1E5     & PN\\
        $-$              &  01(2) &  21\,29\,58.42 &  $+$51\,03\,59.80 & 05602477  &         &         & \\
        $-$              &  01(3) &  21\,29\,58.42 &  $+$51\,03\,59.80 & 15901777  &         &         & \\
        $-$              &  01(2) &  21\,29\,58.42 &  $+$51\,03\,59.80 & 36801940  &         &         & \\
\object{IRAS 22308+5812} &  01(2) &  22\,32\,45.95 &  $+$58\,28\,21.00 & 17701258  & O7.5    & 3E3     & H~{\sc ii}\\
\object{IRAS 23030+5958} &  01(2) &  23\,05\,10.57 &  $+$60\,14\,40.60 & 22000961  & O6.5    & 7E3     & H~{\sc ii}\\
\object{IRAS 23133+6050} &  01(2) &  23\,15\,31.44 &  $+$61\,07\,08.51 & 22001506  & O9.5    & 7E5     & H~{\sc ii}\\
\hline
\hline
\end{tabular}
\\
$^{a}$ SWS observing mode used
\citep[see][]{1996A&A...315L..49D}. Numbers in brackets correspond to 
the scanning speed.\\ 
$^{b}$TDT number which uniquely identifies each ISO observation.\\
$^{\dagger}$These spectra have been obtained by co-adding the separate
SWS spectra also listed in the table, see text.\\
$^{\ddagger}$Effective temperature from \citet{2000A&A...358.1058G}. 
$^{\diamond}$Effective temperature from \citet{2000ApJ...539..783L}.
\end{table*}
We include 29 high S/N spectra of 16 sources taken from the samples of
H~{\sc ii} regions, YSOs, and post-AGB objects. For the trend analysis
presented in Sect.~\ref{SecTrends} we have added another 15 sources in
which the S/N is such that we cannot reliably measure the weakest
features but we determine the intensities of the stronger UIR bands
with reasonable accuracy.  All spectra presented here have been
obtained with the SWS instrument in the AOT01/AOT06 scanning mode at
various speeds, with resolving power ($\lambda/\Delta\lambda$) ranging
from 500$-$1500 \citep[see][]{1996A&A...315L..49D}, see
Table~\ref{tab:obslog} for details of the observations.

The data were processed using SWS interactive analysis product; IA$^3$
\citep[see][]{1996A&A...315L..49D} using calibration files and
procedures equivalent to pipeline version 7.0. If a source has been
observed multiple times and these observations are of similar quality
and of comparable flux-level these data are co-added after the
pipeline reduction. Since the features we discuss here are fully
resolved in all observing modes, we combine the data obtained in all
different modes to maximise the S/N. Further data processing consisted
of bad data removal, rebinning on a fixed resolution wavelength grid,
removing fringes and splicing of the sub-bands to form a continuous
spectrum.

For all spectra the amount of shifting between sub-bands required
falls well within the calibration uncertainties in the region of
interest: 7 to 16 $\mu$m. Any jumps between bands are due to either
flux calibration and dark current uncertainties. The effect of dark
current is most important in low flux cases while the flux calibration
uncertainties will dominate in bright sources. Below 20 Jy we apply
offsets to correct for dark current uncertainties.  In these low
signal cases the typical noise level in the dark current measurements
of 1$-$2 Jy introduces offset uncertainties $>$5$-$10 per cent
dominating the flux calibration uncertainties. Above 20 Jy we apply
scaling factors to correct for flux calibration uncertainties. The
splicing introduces little uncertainty in the measured strengths since
most features fall completely within one ISO/SWS sub-band. An
exception to this is the band strength of the 12.7 $\mu$m feature.
This feature is sensitive to the way band 2C (7 to 12.5 $\mu$m) and 3A
(12 to 16.5 $\mu$m) are combined. This introduces an extra uncertainty
of the band strength of typically 20$-$30 per cent for the weakest
features.

Some SWS data, especially in band 3A, are affected by fringes. We have
corrected for fringes in those sources where they occur, using the
aarfringe tool of IA$^3$ on the rebinned spectrum.  In the method we
apply fringes are fitted with sine functions with periods in the range
where fringes are known to occur and divided out. Note that the
features we study here are much broader than any of fringe periods,
therefore the intensities we measure are not directly affected by the
fringes. However in some cases after fringe removal the continuum is
more easily determined.

SWS spectra of many sources, including stars enshrouded in both
carbon-rich and oxygen-rich dust and sources without any circumstellar
material show very weak structure around 13.5 and 14.2 $\mu$m at the 3
to 4 per cent level relative to the continuum possibly due to residual
instrumental response. The emission features discussed here are all
stronger than this with a maximum of 85 per cent of the continuum in
the reflection nebula NGC 7023. Near 11.03 $\mu$m there is a residual
instrumental feature which coincides with the weak 11.0 $\mu$m feature
that we observe in our spectra. We have included the effect of this
feature in the uncertainty on the intensities in
Table~\ref{TabFeatureProperties}.

Many sources in this sample have strong narrow emission lines in their
spectrum, in particular the strong [Ne~{\sc ii}] line at 12.81 $\mu$m
is perched on top of the 12.7 $\mu$m UIR band. This line and the UIR
band are easily separated at the resolution of the SWS instrument.  We
remove the contribution from this line by fitting a Gaussian profile
to the line and subtracting that profile prior to rebinning.  The
spectrum of NGC 7027 has a very strong [Ne V] emission line at 14.32
$\mu$m. We have removed the part of the spectrum which contains this
line.
  
We also include in Table~\ref{tab:obslog} the spectral type of the
illuminating source and an estimate of the flux density at the
location where the PAH emission originates from in units of the
average interstellar UV field \citep{1968BAN....19..421H}.  We have
derived these estimates from the observed IR flux ($I_\mathrm{IR}$)
and the angular size of the PAH emission region
\citep{1989ApJ...344..770W}. This estimate is based on the assumption
that all the UV light is absorbed in a spherical shell with the
angular size of the object and re-emitted in the IR. The flux density
at the shell is given by:
\begin{equation}
        G_\mathrm{0} = \frac{4}{1.6\,10^{-6}}\left(\frac{1~pc}{1~AU}\right)^2
        \frac{I_\mathrm{IR}}{\theta^2} = 1.1\,10^{17}\,{\theta}^{-2},
\end{equation}
where $G_\mathrm{0}$ is the UV flux density in 1.6\,10$^{-6}$
Watt/m$^2$ and $\theta$ is the angular diameter of the object in
arcseconds.  We have used for the size of the H~{\sc ii} regions the
measured radio sizes.  This is reasonable since the PAHs are expected
to be destroyed inside the H~{\sc ii} region.  For the other objects
we estimate the size of the PAH emission region from ISOCAM data
\citep{1996A&A...315L..32C} when available.

\section{Results}
\label{SecResults}
\begin{figure}
  \psfig{figure=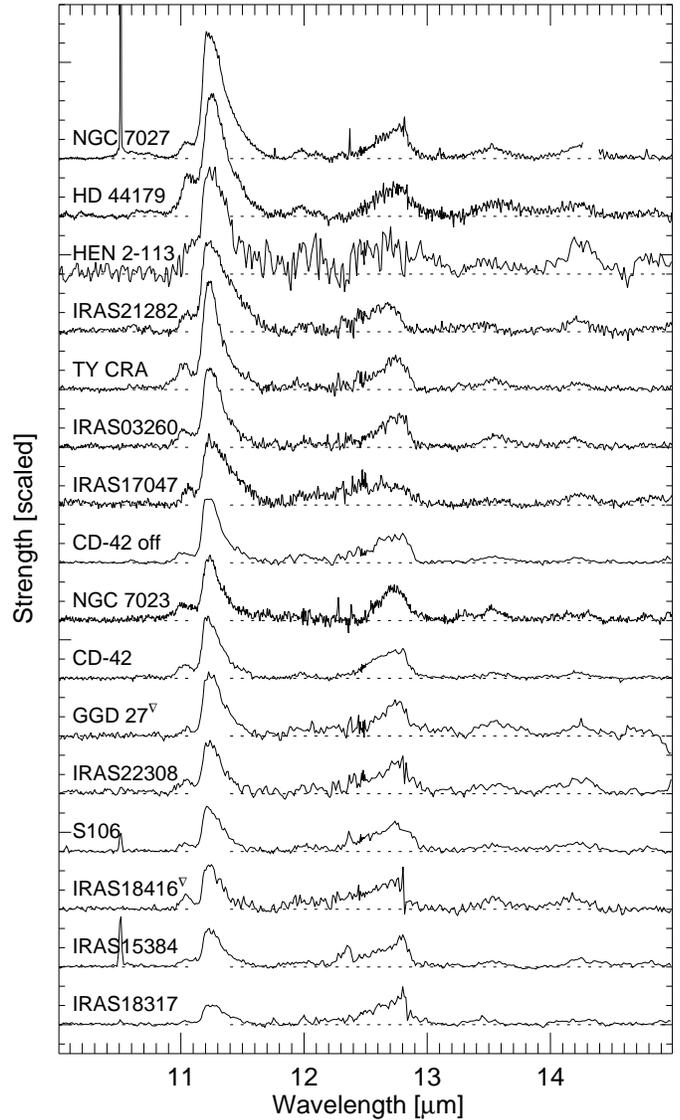,angle=0} 
  \caption{An overview of the observed
    features. All spectra have been continuum subtracted and are scaled
    to have the same integrated intensity in the 12.7 $\mu$m feature.
    The sources are ordered according to their 11.2/12.7 $\mu$m band
    strength ratio (bottom to top). The ratio of the 11.2 $\mu$m to the
    12.7 $\mu$m feature spans a full decade.  \newline$^\nabla$ Sources
    with broad solid CO$_\mathrm{2}$ absorption beyond 15 $\mu$m.}
\label{FigTo127}
\end{figure}
\begin{figure}
  \psfig{figure=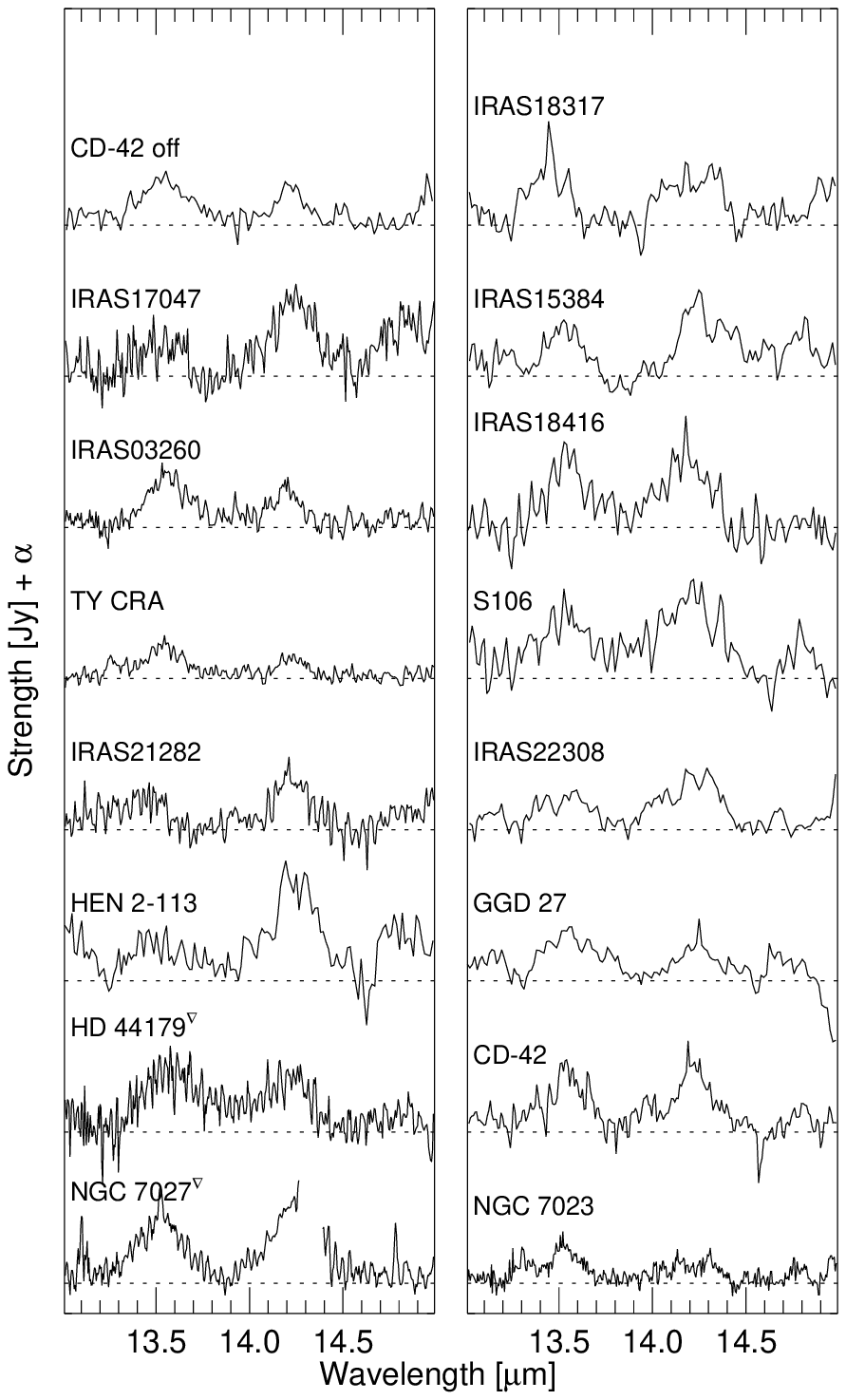,angle=0}
  \caption{An overview of the observed features near 13.5 and 14.2
    $\mu$m. $^\nabla$HD 44179 and NGC 7027 have been scaled by a
    factor of 0.3.\newline}
  \label{FigWeakFeatures}
\end{figure}

\subsection{Overview}
In Fig.~\ref{FigFull} we present spectra of three typical sources to
illustrate the spectral detail present and the variety of underlying
continua. Even more so than previous IRAS/LRS and ground-based studies
suggested \citep{1985ApJ...299L..93C, 1989MNRAS.236..485R,
  1989ApJ...341..270W}, the complete 11$-$15 $\mu$m spectra reveal an
extremely rich collection of emission features with bands at 10.6,
11.0, 11.23, 12.0, 12.7, 13.5, and 14.2 $\mu$m. These features are
perched on top of an emission plateau of variable strength, which
extends across the entire region.

\subsection{Continuum}
In order to compare the intrinsic strength of the UIR bands in the
different sources, we subtract a continuum, splined through points
from 9$-$10.5 and 14.5$-$15.5 $\mu$m and through points near 11.8 and
13.1 $\mu$m. Since GGD 27 and IRAS 18416 have broad solid
CO$_\mathrm{2}$ absorption features beyond 15 $\mu$m, we take 14.7
$\mu$m to be the continuum rather than extending to 15.5 $\mu$m.
Typical examples of the continua are included in Fig.~\ref{FigFull}.
Besides the chosen points, the continuum also runs through the local
minimum at 10.9, 12.2 and 14.0 $\mu$m with the exception of the Red
Rectangle. This object has excess emission around these wavelengths,
which could be related to the presence of crystalline silicates in the
vicinity of HD 44179 \citep{1998Natur.391..868W}. For any reasonable
choice of continuum points, the underlying plateau ranges from about
10 to 14 $\mu$m, peaking at about 12 $\mu$m. At first sight, the
plateau of GGD 27 seems to differ with an onset at about 11 $\mu$m and
peaking around 13 $\mu$m. However, this likely reflects the effects of
foreground silicate absorption. The PN IRAS 21282+5050 does show a
plateau that differs, peaking at 12 $\mu$m but extending all the way
to 17 $\mu$m.

\subsection{Emission bands}
\begin{table*}
  \caption{
    Peak position and strength of the features observed in the
    10$-$15 $\mu$m for the sources shown in
    Fig.~\ref{FigTo127}.} 
  \begin{tabular}{l r r r r r r r r@{\ } c@{\ } c}
    \hline \hline
    \multicolumn{1}{c}{(1)}& 
    \multicolumn{1}{c}{(2)}&
    \multicolumn{1}{c}{(3)}& 
    \multicolumn{1}{c}{(4)}&
    \multicolumn{1}{c}{(5)}& 
    \multicolumn{1}{c}{(6)}&
    \multicolumn{1}{c}{(7)}& 
    \multicolumn{1}{c}{(8)}&
    \multicolumn{1}{c}{(9)}& 
    \multicolumn{1}{c}{(10)}&
    \multicolumn{1}{c}{(11)}\\
    $ Source $ & 
    $ {\Delta\lambda_\mathrm{11.2}}^\dagger$ & $ {I_\mathrm{11.2}}^b$ & 
    $ {I_\mathrm{12.7}}^{b\ddagger}$ & 
    $ {\lambda_\mathrm{c,13.5}}^a$ & $ {I_\mathrm{13.5}}^b$ & 
    $ {\lambda_\mathrm{c,14.2}}^a$ & $ {I_\mathrm{14.2}}^b$ &
    $ {I_\mathrm{11.0}}^{b}$ & $ {I_\mathrm{12.0}}^{b\star}$ & $ {I_\mathrm{plateau}}^{b\diamond}$ \\
    & 
    $ [10^{-3}\mu m] $ & $$ & 
    $$ & $[\mu m]$ &
    $$ & $[\mu m]$ & 
    $$ & $$ &
    $$& $$ \\
    \hline 
    IRAS 03260 &   0.9(0.1) &  15.8(0.4) &   6.4(0.4) & 13.57(2) &  1.7(0.1) & 14.19(2) &  0.9(0.1) &  0.8(0.1) &n &  22 \\
    HD 44179 &  12.6(0.5) & 118.1(6.4) &  26.3(1.4) & 13.61(2) & 13.4(0.5) & 14.21(2) &  8.6(1.3) &  4.8(2.3) &5 & 112 \\
    HEN 2-113 &   2.7(0.3) &  17.0(1.0) &   5.6(1.0) &  $-$     &  $-$      & 14.26(5) &  4.7(1.8) &  0.5(0.5) &d &  30 \\
    IRAS 15384 &   0.0(0.6) &  14.5(0.8) &  13.5(0.6) & 13.52(2) &  1.5(0.3) & 14.30(2) &  2.7(0.1) &  0.6(0.1) &2 &  66 \\
    CD -42(off)&  -6.5(0.2) &  22.2(0.6) &  11.3(0.5) & 13.55(2) &  1.7(0.5) & 14.22(2) &  0.8(0.2) &  1.0(0.1) &2 &  25 \\
    CD -42 &  -8.2(0.8) &  31.7(1.5) &  16.7(1.0) & 13.53(2) &  1.6(0.6) & 14.22(2) &  1.6(0.3) &  1.9(0.3) &1 &  58 \\
    IRAS 17047 &  11.8(0.4) &  25.1(1.0) &  11.6(1.0) & 13.50(3) &  1.8(1.3) & 14.23(3) &  2.2(1.0) &  1.3(0.8) &n & 145 \\
    GGD 27 &   2.6(0.6) &   8.9(0.6) &   5.4(0.5) & 13.57(2) &  2.1(0.2) & 14.27(4) &  1.5(0.9) &  1.4(0.1) &d &   3 \\
    IRAS 18317 &   6.5(0.1) &  10.5(0.2) &  15.6(0.3) & 13.44(2) &  2.0(0.3) & 14.22(2) &  2.2(0.2) &  0.2(0.1) &1 &  83 \\
    IRAS 18416 &  -2.3(0.3) &  12.8(0.9) &   9.8(0.5) & 13.55(2) &  2.4(0.4) & 14.18(3) &  3.1(1.2) &  1.2(0.3) &n &  52 \\
    TY CRA &  -5.0(0.3) &  15.5(0.1) &   4.7(0.2) & 13.54(2) &  1.3(0.3) & 14.21(2) &  0.5(0.1) &  0.8(0.1) &n &  13 \\
    S 106 &  -4.1(0.1) &  19.7(0.9) &  15.6(1.0) & 13.56(2) &  2.7(1.2) & 14.19(2) &  3.6(0.2) &  1.1(0.3) &2 &  61 \\
    NGC 7023 &  -3.9(0.2) &   9.6(0.8) &   4.3(0.3) & 13.50(2) &  1.1(0.2) & 14.21(2) &  0.7(0.3) &  0.5(0.1) &n &  11 \\
    NGC 7027 &   2.8(0.2) & 142.7(5.4) &  35.9(2.5) & 13.52(2) &  9.7(1.5) & 14.26(2) & 11.1(1.5) &  4.0(1.7) &6 & 451 \\
    IRAS 21282 &   4.2(1.4) &  20.4(0.7) &   6.2(0.2) & 13.40(2) &  1.5(0.4) & 14.22(2) &  1.7(0.6) &  0.6(0.3) &y & 104 \\
    IRAS 22308 &  -3.2(0.5) &   8.9(0.4) &   5.8(0.6) & 13.55(4) &  1.3(0.6) & 14.23(2) &  2.1(0.1) &  0.7(0.1) &n &   9 \\
    \hline
    \hline
  \end{tabular}\\
  The intensities $I_\mathrm{11.2}$,
  $I_\mathrm{12.7}$, $I_\mathrm{11.0}$, $I_\mathrm{12.0}$ and
  $I_\mathrm{plateau}$ are the 
  integrated fluxes of the features after continuum subtraction
  (columns 3,4,9,10 and 11). The position of the 11.2 and 12.7
  $\mu$m features are determined by fitting the template profile to
  the spectra in which we allow for a wavelength shift and a scaling
  of the template (column 2). The properties of the features around
  13.5 and 14.2 are determined by fitting Gaussian profiles to the data
  (columns 5$-$8). Numbers in parentheses are uncertainties.\\
  $^a$Central wavelength of the fitted Gaussian profile, uncertainties
  are given in parentheses in units of 10$^{-2}$$\mu$m.\\ 
  $^b$Intensities and uncertainty units are 10$^{-14}$W/m$^2$\\ 
  $^\dagger$Shift of the 11.2 $\mu$m feature relative to the template
  11.2 $\mu$m profile. The template profile peaks at 11.229(0.001)
  $\mu$m.\\
  $^{\ddagger}$No significant shifts of the 12.7 $\mu$m profile are
  observed with respect to the mean 12.7 $\mu$m profile. The mean
  profile ranges from 12.2$-$12.95 $\mu$m with the peak position at
  12.804(0.005) $\mu$m. Typical uncertainty in the position
  determination 0.006 $\mu$m.\\
  $^{\star}$Typical error on $I_\mathrm{12.0}$ is
  0.5$\cdot$10$^{-14}$W/m$^2$; d means detected but not measured; n
  means not detected.\\
  $^{\diamond}$Typical error on $I_\mathrm{plateau}$ is
  10$\cdot$10$^{-14}$W/m$^2$.\\
  \label{TabFeatureProperties}
\end{table*}
We show the continuum subtracted spectra of the 16 objects with the
highest S/N in Fig.~\ref{FigTo127}. Perusal of the spectra reveals a
plethora of weaker features. NGC 7027, HD 44179 and IRAS 21282 have a
very weak, broad feature near 10.6 $\mu$m. Almost all these sources
show a feature at 11.0 $\mu$m, except for GGD 27, where the feature is
possibly present but only at the 1$\sigma$ level. The weak 12.0 $\mu$m
band is detected in 10 out of 16 sources. The 13$-$15 $\mu$m range
contains the newly discovered weak 13.5 and 14.2 $\mu$m features
(Fig.~\ref{FigWeakFeatures}). Structure near 13.5 is present in all
sources except HEN 2-113, however for IRAS 21282 and IRAS 18317 the
band is replaced by a feature that peaks at a shorter wavelength.  A
feature near 14.2 $\mu$m is found in all sources although the actual
peak position varies considerably (cf., IRAS 03260 and IRAS 15384).
This variation in peak position of these two weak bands is quite in
contrast with that of the 11.2 and 12.7 $\mu$m bands which almost
invariably peak at about the same wavelength (see below). \emph{A
  priori}, it is not given that the spectral structure near 14 $\mu$m
is actually related in the different sources.  Possibly, rather than
one molecular vibration with a varying peak position, at this weak
level of emission, we are probing different bands whose relative
strengths reflect the conditions in the different sources.

In Fig.~\ref{FigTo127} we show the continuum subtracted spectra after
normalising to the integrated strength of the 12.7 $\mu$m feature. The
sources are ordered according to the strength of the 11.2 $\mu$m
feature relative to the 12.7 $\mu$m band. Relative to the 12.7 $\mu$m
band, the sources with the weakest 11.2 $\mu$m feature are the H~{\sc
  ii} regions (at the bottom of Fig.~\ref{FigTo127}), while the
evolved stars show the strongest 11.2 $\mu$m band.

The spectral characteristics of the features are summarised in
Table~\ref{TabFeatureProperties}. Note that the uncertainties quoted
in the Table reflect the noise level and the freedom in drawing the
continuum \emph{within} the methodology used to measure these bands.
Other ways of decomposing the broad, blended bands and the underlying
continuum will give other results
\citep[e.g.][]{1998A&A...339..194B,2000ApJ...530..817U,2001Verstraete}.
However these differences are \emph{systematic} differences and do not
affect the source-to-source variations we observe. The intensities of
the 11.2 and 12.7 $\mu$m features are obtained by direct integration
above the chosen continuum. We measure the peak position of the 11.2
and 12.7 $\mu$m bands by fitting them with template spectra of these
features. The template spectrum for the 11.2(12.7) $\mu$m feature is
constructed by adding the continuum subtracted spectra with each
11.2(12.7) $\mu$m feature normalised to have the same integrated
intensity. This way each source has equal contribution to the template
spectrum. We use a $\chi^2$-minimisation routine to fit the template
to the sources, allowing for both a wavelength shift and scaling in
strength. The shifts that we determine for the 11.2 $\mu$m band are
very small except for HD 44179 and IRAS17047 where this band is much
broader than the template spectrum (cf.,
Table~\ref{TabFeatureProperties}, see also Peeters et al. (2001, in
prep.)).  Although there are differences between the detailed profiles
of the 12.7 $\mu$m band we detect no significant shift of the band as
a whole. For the weak features near 13.5 and 14.2 $\mu$m, the
parameters have been determined through fitting of Gaussian profiles.
We adopted a local linear continuum for the very weak 11.0 $\mu$m
feature because of the severe blending of this band with the 11.2
$\mu$m band. The weak 12.0 $\mu$m band is close to both the 11.2 and
the 12.7 $\mu$m band. For only a few sources we measure the intensity
of this band, for the other sources we refrained from detailed
analysis. However Table~\ref{TabFeatureProperties} does note whether
we detect this band.

The profile of the 11.2 $\mu$m feature is asymmetric with a sharp blue
rise and a more gradual decline to longer wavelengths
\citep{1989MNRAS.236..485R,1989ApJ...341..270W}. This will be
discussed in more detail for this sample by Peeters et al. (2001, in
prep.). The 12.7 $\mu$m band is also asymmetric but in the opposite
way with a slow blue rise and a sharp red decline between 12.8 and
12.9 $\mu$m.  Because of their intrinsic weakness, the profiles of the
10.6, 11.0, 12.0, 13.5, and 14.2 $\mu$m features in the individual
sources are not well determined however in the averaged spectrum,
these features appear symmetric (cf., Fig.~\ref{FigLabBands}).

\subsection{Trends}
\label{SecTrends}
Given the large range of band strength ratios in the region of
interest and the smooth variations in this ratio, it is of interest to
investigate the correlation of the strength of all UIR emission bands
with these variations. As a class all the planetary nebulae with a
Wolf-Rayet central star (HEN 2-113, BD+30, IRAS17047, IRAS 07027 and
IRAS 18240) as well as the post-AGB star HD 44179 show a distinctly
different UIR spectrum characterised by a shift in of the 6.2 and 7.7
$\mu$m bands towards 6.3 and 8 $\mu$m respectively (Peeters et al.,
2001, in prep.).  Because of these spectral differences we have not
included them in the following trend analysis in which they also
separate out as a peculiar subgroup. These objects will be discussed
in detail in a forthcoming paper (Hony et al., 2001, in prep.).

\begin{figure}
  \psfig{figure=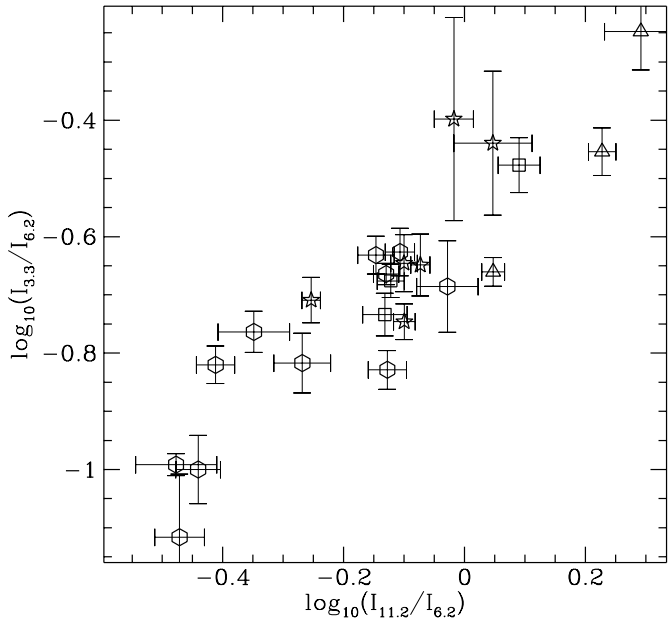,width=8.8cm}
  \caption{Bands strength ratios as derived from the SWS
    spectra. Hexagons are H~{\sc ii} regions, stars intermediate mass
    star forming regions, squares RNe and triangles are
    PNe.}
  \label{FigTrend1}
\end{figure}

\begin{figure}
  \psfig{figure=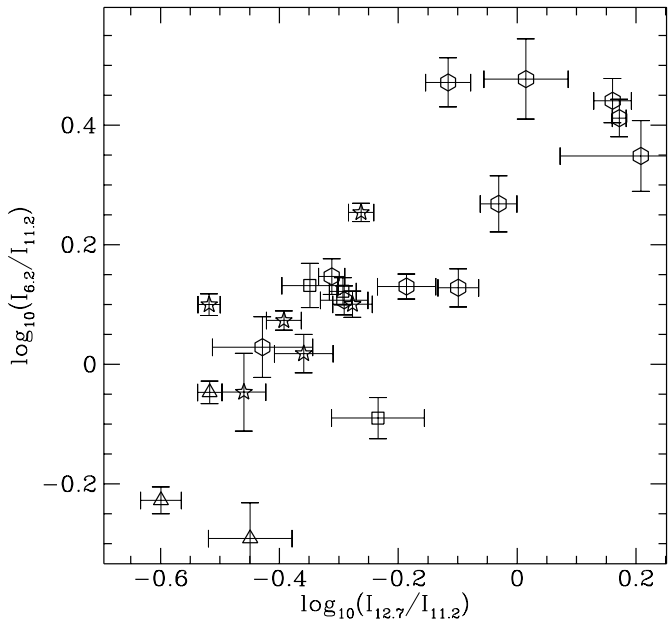,width=8.8cm}
  \caption{Bands strength ratios as derived from the SWS
    spectra. Plotting symbols are the same as in
    Fig.~\ref{FigTrend1}.} 
  \label{FigTrend2}
\end{figure}

\begin{figure}
  \psfig{figure=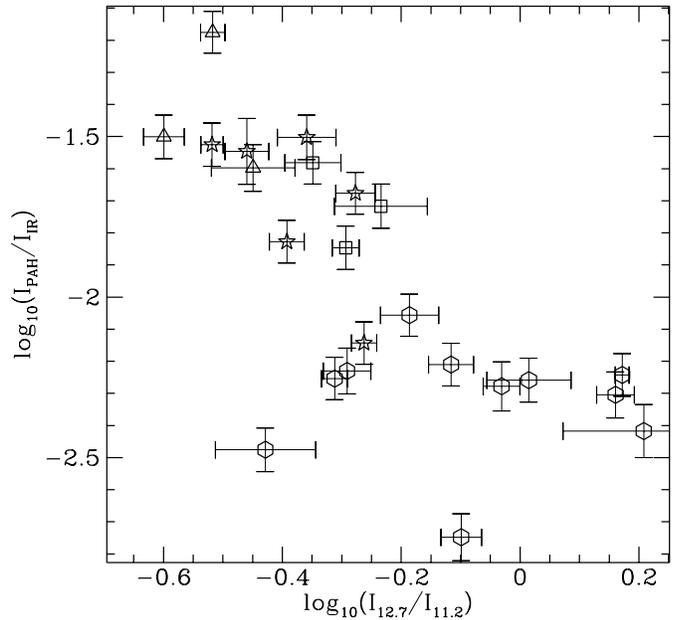,width=8.8cm}
   \caption{The ratio between the flux emitted in the PAH bands over
     the total amount of IR radiation against the
     $I_\mathrm{12.7}$/$I_\mathrm{11.2}$ ratio. Plotting symbols are
     the same as in Fig.~\ref{FigTrend1}.}
  \label{FigTrend5}
\end{figure}

Because here we want to study variations in the \emph{relative}
strength of the UIR bands to each other, not differences in absolute
intensities differences due to intrinsic luminosity and distance of
the source, we use 3-feature intensity ratio correlations.  Although
we observe variations in all ratios, we find only three that correlate
and these are shown in Figures~\ref{FigTrend1}$-$\ref{FigTrend5}.
First, we find that the CH stretch mode at 3.3 $\mu$m correlates with
the 11.2 $\mu$m band (cf.  Fig.~\ref{FigTrend1}). Note that the slope
of the trend is roughly 1, which means that the
$I_\mathrm{11.2}$/$I_\mathrm{3.3}$ is on the average constant at a
value of 3$-$4.

Second, the 12.7 $\mu$m band correlates with the CC stretch mode at
6.2 $\mu$m, albeit with more scatter (cf. Fig.~\ref{FigTrend2}) than
the 11.2/3.3 ratio. The 7$-$9 $\mu$m complex is well correlated with
the 6.2 $\mu$m band and similar plots can be made with these
interchanged. Inspection of Fig.~\ref{FigTo127} and
Table~\ref{TabFeatureProperties} shows that there is some indication
for both the $I_\mathrm{13.5}$/$I_\mathrm{11.2}$ and the
$I_\mathrm{14.2}$/$I_\mathrm{11.2}$ $\mu$m band to be higher in H~{\sc
  ii} regions, however only about half the sources have such high S/N
that these intensities can be reliably measured and this trend is not
statistically significant.

Lastly, we show in Fig.~\ref{FigTrend5} the correlation between the
ratio of the flux emitted in the PAH bands relative to the total flux
emitted in the IR ($I_\mathrm{IR}$) and the changing
$I_\mathrm{12.7}$/$I_\mathrm{11.2}$ ratio.  We measure the
$I_\mathrm{IR}$ by integrating the SWS data and Long Wavelength
Spectrometer (LWS) data if available. For those sources without LWS
data we use a blackbody fitted to IRAS measurements in the wavelength
region from 45$-$200 $\mu$m. We do not apply corrections for aperture
differences between the instruments. We estimate an uncertainty of 15
per cent on the $I_\mathrm{IR}$. Again different classes of objects
occupy different parts in this diagram.

We also checked for correlations between band strength ratios and the
flux density; $G_\mathrm{0}$, however we do not detect any
correlations.

We emphasise that, while all UIR bands show a loose correlation in the
absolute intensity \citep[see also][]{1986ApJ...302..737C,
  1989ApJ...341..246C}, these three are the only tight correlations
present in this sample.

\subsection{Comparison with other studies}
It is interesting to compare the result we obtained from the SWS
spectra with results by other authors. Studying many locations in the
diffuse interstellar medium \citet{2000ibp..confE..14C} find that the
relative bands strengths of the 6.2, 7.7 and the 11.2 $\mu$m features
do not vary systematically over a wide range of intensities of the
incident radiation field. Their observed ratios of
$I_\mathrm{11.2}$/$I_\mathrm{6.2}$ cluster around 0.8, the value we
observe in the RNe and the YSOs. Studying a few RNe at various
locations \citet{2000ApJ...530..817U} with ISO/CAM
\citep{1996A&A...315L..32C} find only small variations in band
strength ratios. The $I_\mathrm{11.2}$/$I_\mathrm{6.2}$ ratios they
derive with the method which is most similar to ours (method 2) are
like those we find for the RNe.  Their
$I_\mathrm{12.7}$/$I_\mathrm{11.2}$ ratios are however systematically
lower that ours. This is due to the lower spectral resolution of the
CAM spectra which results in blending and smearing of the weakish band
at 12.7 $\mu$m.  The SWS spectra are not affected by smearing since
the features are fully resolved. The results of these authors, like
the results we present here, demonstrate that similar types of sources
show similar PAH band strength ratios. Our results also demonstrate
how \emph{different classes of objects show systematic differences in
  their PAH spectra}.
  
Observations of the starforming region M17 have suggested that the
13.5 $\mu$m band is correlated with the mid-IR continuum
\citet{1996A&A...315L.337V}. Such a correlation is important to
establish since it might yield information on the size of the carriers
of the 13.5 $\mu$m band. We have examined whether a correlation is
also present in the sample of sources we study here.  We have
therefore compared the 15$-$16 $\mu$m continuum with the strength of
the 13.5 $\mu$m band. We find strong variations, by a factor of
$\simeq$100 in the strength of the mid-IR continuum relative to the
13.5 $\mu$m band. These strong variations are not surprising
considering the fact that we look at very diverse regions with large
differences in dust composition and temperature distributions. For
example the evolved object HD44179 has contributions from crystalline
silicates around 15 $\mu$m \citep{1998Natur.391..868W}.  Many H~{\sc
  ii} regions have a strongly rising continuum due to warm dust.
However we would also like to point to the two observations of CD-42
11721 where the 13.5 band is equally strong but the dusty continuum is
missing in the off-pointed observation. These two observations show
the 15 micron continuum and the 13.5 micron feature to be decoupled
even within the same object.

\section{The CH out$-$of$-$plane bending modes}
\begin{table}
  \caption{
    Wavelength region limits and the integrated absorption
    cross-sections for the CH out$-$of$-$plane bending modes.}
  \begin{tabular}{@{\,}l@{\,} r@{\,} r@{\,} r@{\,} r@{\,} r@{\,}}
    \hline
    \hline
    \multicolumn{1}{c}{$$}& 
    \multicolumn{1}{c}{${\lambda_\mathrm{low}}^\dagger$}&
    \multicolumn{1}{c}{${\lambda_\mathrm{up}}^\dagger$}& 
    \multicolumn{1}{c}{$A^\diamond\ddagger$}&
    \multicolumn{1}{c}{${A_\mathrm{neutral}}^\diamond$}&
    \multicolumn{1}{c}{${A_\mathrm{cation}}^\diamond$}
    \\
    &
    \multicolumn{1}{c}{$[\mu m]$}&
    \multicolumn{1}{c}{$[\mu m]$}& 
    \multicolumn{1}{c}{$[km/mol]$}&
    \multicolumn{1}{c}{$[km/mol]$}&
    \multicolumn{1}{c}{$[km/mol]$}
    \\
    \hline
    Solo    & 10.6  & 11.4 & 24.8(13.5) & 25.7(14.2)& 24.1(12.9)\\
    Duo     & 11.35 & 12.8 &  4(2.5)    &  4.4(2.4) & 3.7(2.5)\\
    Trio    & 12.5  & 13.3 &  9.6(5.9)  & 10.1(5.3) & 9.0(6.5) \\
    Quartet & 13.0  & 13.9 & 12.0(4.8)  & 11.5(5.5) & 12.6(3.9)\\
    \hline
    \hline
  \end{tabular}
  \newline
  Summary of the laboratory results on CH out$-$of$-$plane bending
  modes for solo, duo, trio and quartet hydrogens on matrix isolated
  neutral polycyclic aromatic hydrocarbons and their cations. (Adapted from
  \citet{2000Hudgins}).\\
  $^\dagger\lambda_\mathrm{low}$ is the lower limit of the region in
  $\mu$m. $\lambda_\mathrm{up}$ is the upper limit of the region in
  $\mu$m. \\
  $^\diamond$The cross-section values for the solo, trio, and quartet
  modes per hydrogen are the averages over the spectra in the database.
  However, the A values for the duo mode per hydrogen decreases rapidly
  with size and settles to slightly less than 4 km/mole for PAHs with
  more than 24 carbon atoms. This value is more appropriate to use in
  determining the edge structures of PAHs that dominate emission in
  this wavelength region.\\
$^\ddagger$Total average cross-sections over both neutrals and cations in
    the database.
\label{TabLabBands}
\end{table}
\begin{figure}
  \psfig{figure=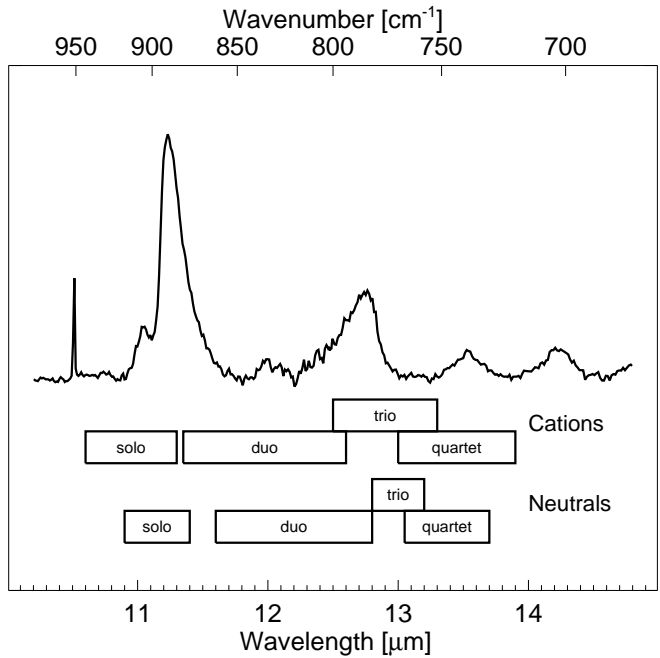,angle=0}
  \caption{
    A comparison of the average interstellar spectrum (top) with the
    ranges for the out$-$of$-$plane bending modes (bottom). The
    average spectrum was obtained by co-adding the continuum
    subtracted spectra after normalisation to the 12.7 $\mu$m band
    strength.  The boxes indicate the wavelength regions associated
    with the out$-$of$-$plane bending vibrations for different types
    of adjacent hydrogen atoms determined from matrix isolated
    spectroscopy of neutral and cationic PAHs (see \citet{2000Hudgins}
    for details). In this comparison it should be kept in mind that
    the emission process leads to a small ($\simeq$ 0.1 $\mu$m)
    wavelength redshift in the peak position.}
  \label{FigLabBands}
\end{figure}

\subsection{Laboratory spectroscopy of the OOP modes}
\label{SecLabdata}
Chemists have long recognised the diagnostic value of the aromatic CH
out$-$of$-$plane bending features in the 11 to 15 $\mu$m spectral
region for the classification of the aromatic ring edge structures
present in a particular sample \citep[e.g.][]{1958Bellamy}.
Specifically, the positions of the bands in this spectral region
reflect the number of adjacent CH groups on the peripheral rings of
the PAH structure \citep{1958Bellamy,ATB85, 1985ApJ...299L..93C,
  1989A&A...216..148L, 1989MNRAS.236..485R, 1989ApJ...341..270W,
  1999ApJ...511L.115A, 1999ApJ...516L..41H}.  Traditionally, aromatic
rings carrying CH groups which have no neighbouring CH groups (termed
"non-adjacent" or "solo" CH groups) show IR activity between 11.1 and
11.6 $\mu$m.  Likewise, activity between 11.6 and 12.5 $\mu$m is
indicative of two adjacent CH groups ("doubly-adjacent" or "duet"
CH's) on the periphery of the PAH. Three adjacent CH groups
("triply-adjacent" or "trio" CH's) are indicated by activity in the
12.4 to 13.3 $\mu$m region, and four adjacent CH groups
("quadruply-adjacent" or "quartet" CH's) by activity between 13 and
13.6 $\mu$m. Five adjacent CH groups ("quintuply-adjacent" or quintet
CH's) are indicated by features falling in the 13 to 13.7 $\mu$m
range. Trios and quintets also show a weak CCC bending mode in the
14$-$14.5 range. Other CCC bending modes occur in the 15$-$20 $\mu$m
range and have been discussed in the astrophysical context by
\citet{2000A&A...357.1013V} and \citet{2000A&A...354L..17M}. Over the
years the reliability of this region to yield insight into the
molecular structure and ring sidegroup placement on aromatic samples
has been verified again and again (see \citet{1999ApJ...516L..41H} and
references therein). However, most of these chemist's guidelines were
based on studies of small PAHs where varying patterns of sidegroup
substitution were employed to achieve different degrees of CH
adjacency.  Furthermore, these chemist's `rules-of-thumb' are based on
spectroscopic studies of aromatic molecules in solution or solid
mixtures, environments quite different from that of the emitting
aromatic species giving rise to the interstellar features presented
here. To obtain infrared data more relevant to the interstellar
situation on larger molecules with the spectral details in this region
determined by PAH structure rather than side-group substitution
pattern, several groups have undertaken new spectroscopic studies of
PAHs carried out under more appropriate conditions (e.g. see
\citet{1995CPL...245..539S} and references therein;
\citet{2000HudginsB} and references). Thanks to this effort, the
mid-IR spectra of a few gas phase and many matrix isolated neutral and
ionised PAHs are now available. In the Astrochemistry Laboratory at
NASA Ames this work has been expanded considerably, including an
extensive set of theoretical calculations, aimed to specifically
address the questions raised by these new ISO spectra. Since a
detailed presentation and analysis of the 11 to 15 $\mu$m region of
the expanded dataset of matrix isolated PAH spectra will be published
separately \citep{2000Hudgins}, only the salient points are summarised
here and used in the analysis presented below.  The IR spectra of
matrix isolated PAHs compare favourably with the available spectra of
gas phase PAHs, for both neutral PAHs and cationic PAHs \citep[see
also][]{1999ApJ...520L..75P} and validates the use of the matrix
isolation method to obtain astronomically relevant data.

There are two points that emerge from an analysis of the laboratory
database that are of particular importance to the observational data
presented here. The first involves the effect of ionisation on the
characteristic wavelength regions of the various CH adjacency classes.
The second is the intrinsic integrated absorption strengths (A values)
which are derived for the various adjacency classes. Together these
results provide the tools to not only \emph{qualitatively} infer the
sorts of PAH edge structures present, but also \emph{quantitatively}
determine their relative amounts. As shown below, this allows one to
place stringent constraints on the emitting interstellar PAH family.

\subsection{OOP modes in the interstellar spectrum}
Fig.~\ref{FigLabBands} and Table~\ref{TabLabBands} summarise the key
points presented in \citet{2000Hudgins} that are applicable to this
work. Fig.~\ref{FigLabBands} schematically compares the average UIR
spectrum with the wavelength regions associated with different
hydrogen type for neutral and ionised isolated PAHs, while
Table~\ref{TabLabBands} lists the specific wavelength limits and the
integrated band strengths per CH group as a function of hydrogen
adjacency for all PAHs in the NASA Ames database. The regions
indicated for the neutral PAHs differ slightly from those indicated by
\citet{1958Bellamy}. We deem our results more astrophysically
representative because of the larger set of molecules studied and
because no substitution with strongly electro-negative groups were
involved. Moreover our data were measured on isolated PAHs in inert
matrices, rather than in solid mixtures.

Perusal of Fig.~\ref{FigLabBands} and the wavelength limits listed in
Table~\ref{TabLabBands} shows that, while the ranges for neutral PAHs
are not modified substantially compared to \citet{1958Bellamy},
ionisation causes some important changes in region boundaries. These
data expand on the initial report that the PAH cation solo hydrogen
position is substantially blue shifted with respect to the wavelength
for its neutral counterpart while the domains indicative of the other
types of hydrogen are less affected by ionisation
\citep{1999ApJ...516L..41H}. Considering these modified domains and
taking into account the roughly 0.1 $\mu$m redshift in the peak
position for PAHs emitting at temperatures of $\sim$ 500$-$1000 K
\citep{1991ApJ...380L..43F, 1992ApJ...388L..39B, 1992ApJ...385..577C,
  1995A&A...299..835J, 1998ApJ...493..793C} allows us to draw the
following conclusions.
\begin{itemize}
\item{The broad, weak interstellar emission feature between 10.6 and
    10.7 $\mu$m and the stronger distinct interstellar band peaking
    near 11.0 $\mu$m fall in the region unambiguously attributable to
    PAH cations.}
\item{Adjusting for the 0.1 $\mu$m redshift, the bulk of the 11.2
    $\mu$m interstellar band falls squarely within the solo region for
    the neutral PAHs and also, at the long wavelength end of the range
    for solo, cationic aromatic CH bonds. }
\item{The domains indicated in Fig.~\ref{FigLabBands} show that
    regardless of the region definitions used, there can be little
    doubt that the weak interstellar 12 $\mu$m band arises from duo
    modes.}
\item{Interestingly, the blueward skewed profile of the moderately
    strong band at 12.7 $\mu$m seems to fall better within the
    envelope for trio hydrogens in ionised PAHs and does not agree
    well with the envelope for trio modes in neutral PAHs. An origin
    in duo modes of neutral PAHs is also consistent
    (Fig.~\ref{FigLabBands}).}
\item{The weak 13.5 $\mu$m feature falls in the quartet domain.
    However, this also overlaps the lower of the two domains
    characteristic of the quintet region, and so, although we consider
    this highly unlikely, it is possible that quintet types of PAH
    structure could contribute to this band as well.}
\end{itemize}
\label{SecCompareLabISM}
\begin{figure}
  \psfig{figure=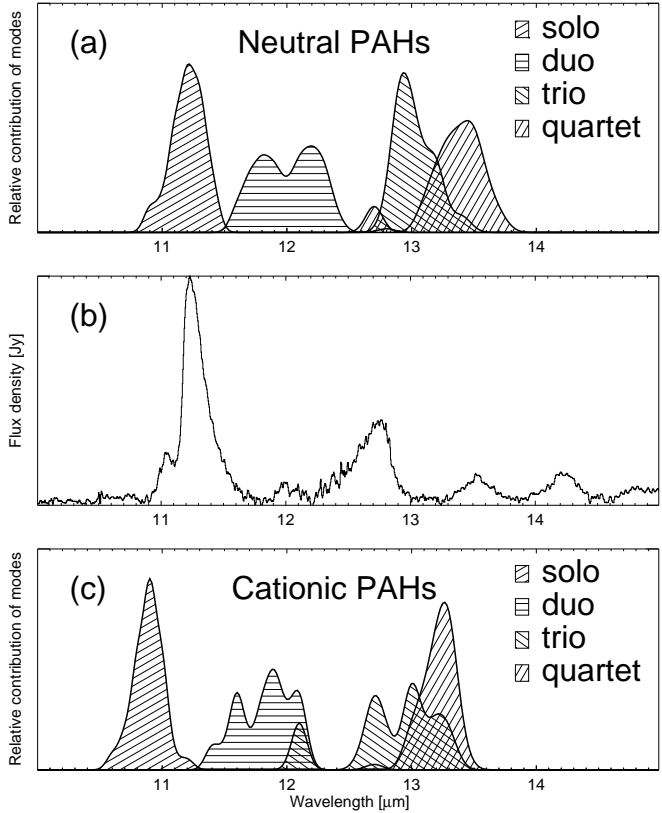}
  \caption{
    Comparison between the mean interstellar spectrum (panel (b)) and
    synthetic PAH spectra showing the distribution over peak positions
    of the OOP modes in the Hudgins database \citep{2000Hudgins}. The
    shaded surfaces in panel (a) and (c) represent the contributions
    per mode for neutral and positively charged PAHs respectively.
    Each area represents the average absorption cross-section per
    functional group.}
  \label{FigLabspec}
\end{figure}
The large laboratory database and the high quality of the ISO
interstellar spectra allows us to compare the \emph{distribution} of
peak positions of the individual modes with the bands observed in the
interstellar spectrum. From the measured peak positions and
cross-sections from the Hudgins database we construct a synthetic
spectrum of a mixture of PAHs for comparison with the observed
interstellar UIR bands. We construct such a spectrum on a
mode-per-mode basis. Per mode, we take for each molecule which shows
this mode the measured peak position and cross-section per CH bond and
convolve this with a Gaussian profile. These contributions we add.
The resulting envelopes per mode are normalised to the number of
molecules exhibiting this mode and corrected for the number of CH
bonds in one functional group. The measured cross-sections for the duo
modes show a systematic decrease with increasing PAH size settling to
a value of $\simeq$4 km/mol. To be consistent with the cross-section
for the largest measured PAHs the contribution of the duo modes have
been scaled down by a factor 1/4. This way each envelope shows the
distribution over wavelengths, while each area corresponds to the
average absorption cross-section per \emph{group} in the database, see
Table~\ref{TabLabBands}.  Thus the synthetic spectrum for each mode is
given by:
\begin{equation}
  S_m(\lambda)=\left(\sum_i
    G(\lambda,A_{m,i},\lambda_{\mathrm{peak},m,i})\right)
  \frac{N_{\mathrm{CH},m}W_m}{N_{\mathrm{mol},m}}, 
\end{equation}
where $S_m$ is the synthetic spectrum for mode $m$, $\lambda$ the
wavelength, $A_{m,i}$ the cross-section of the mode $m$ in molecule
$i$ and $\lambda_{\mathrm{peak},m,i}$ the corresponding peak position.
$G(\lambda,X,Y)$ is a Gauss function with FWHM=10 cm$^{-1}$,
surface=$X$, peak position=$Y$. $N_{\mathrm{CH},m}$ designates the
number of CH bonds in one functional group, $N_{\mathrm{mol},m}$ the
total number of molecules with mode $m$.  $W_m$ is a weighing factor
which equals 1, 1/4, 1 and 1 for solo, duo, trio and quartet modes
respectively to account for the decrease in strength of the duo modes
in larger PAHs. The summation is done over all molecules in the
database. The resulting envelopes are shown in Fig.~\ref{FigLabspec}
for both the neutral and cationic PAHs.

One should bear in mind, that the measured species are probably
smaller than those that dominate the interstellar population and that,
for stability reasons, the interstellar PAH family might be skewed to
a few of these molecules or the edges structures they represent (cf.
Sect.~\ref{SecMolstructure}). Comparing the interstellar spectrum
(Fig.~\ref{FigLabspec}b) with these averaged laboratory spectra
(Fig.~\ref{FigLabspec}a,c) allows us to further refine the discussion
of Sect.~\ref{SecLabdata}:
\begin{itemize}
\item{The position of the strongest band at 11.23 $\mu$m agrees well
    with the measured position of solo transitions in neutral PAHs but
    does not agree with the position of the cationic solos even after
    including a 0.1 $\mu$m shift due to the high temperature of
    interstellar PAHs. It is however important to note that the solo
    mode of the largest cation in the database peaks at 11.2 $\mu$m,
    the longest wavelength of all measured cationic solo modes.}
\item{The peak of the 12.7 $\mu$m emission feature falls at slightly
    shorter wavelength than the centre of weight for both the neutral
    and cationic trio modes. Also, the blue wing of the 12.7 $\mu$m
    band in the interstellar spectra does not coincide with any strong
    emission bands of the PAH species in the database.}
\item{The centre of weight of the neutral quartet vibrations is 13.4
    $\mu$m which matches better with the position of the 13.5 $\mu$m
    UIR band in the interstellar spectrum than does the centre of
    weight for cationic quartets (13.25 $\mu$m).}
\end{itemize}
Summarising the above we find that the overall match between the UIR
spectra and the neutral species is best. There are significant
differences between the combined laboratory measurements and the
interstellar spectra. In particular the precise assignment of the 12.7
$\mu$m feature is uncertain. Given the above observations, we feel
that while the assignment with trio modes is attractive, the case in
not completely compelling. These issues might be resolved when larger
species are measured in the laboratory. However it is clear that the
11.0 and 11.2 $\mu$m feature are due to \emph{solo} CH bonds in
ionised and neutral PAHs respectively, and the bands at longer
wavelength are due to \emph{multiplets}.

It is also immediately clear from Fig.~\ref{FigLabspec} that the
interstellar spectrum does not reflect an equal distribution over the
different functional groups but is dominated by the contribution of
solo modes. This reflects the molecular structure of the emitting
PAHs.

\section{The molecular structures of interstellar PAHs}
\label{SecMolstructure}
\begin{table}
  \caption{
    Relative number of solo,duo,trio and quartet groups in NGC 7027
    and IRAS 18317.}
  \begin{center}
    \begin{tabular}{l c c c}
      \hline
      \hline
      &
      $s/d$& 
      $s/t$&
      $s/q$
      \\
      \hline
      NGC 7207& 7.7 & 4.6 & 28.5\\
      IRAS 18317 & 3.4& 0.8 & 10.2\\
      \hline
      \hline
    \end{tabular}
  \end{center}
  The ratio of the number of solo to duo ($s/d$), solo
  to trio ($s/t$) and solo to quartet($s/q$) groups
  for NGC 7027 and IRAS 18317 as deduced from their 10 to 15 $\mu$m
  spectra.
  \label{TabAstroRatios}
\end{table}
\begin{figure}
  \centerline{
    \psfig{figure=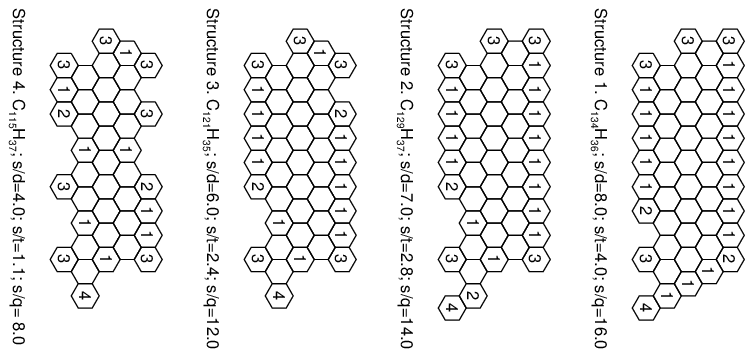,angle=90,height=10cm}}
  \caption{
    \emph{Examples} of molecular structures simultaneously satisfying
    the structural constraints set by the observed band strength
    ratios of the number of solo,duo and trio modes for different
    interstellar regions. The number of solo,duo,trio and quartet
    functional groups are noted $s,d,t$ and $q$ respectively. Solo
    modes are associated with long straight molecular edges. Duos and
    trios, on the other hand, correspond to corners. Quartets are due
    to pendant rings attached to the structure. The numbers in the
    molecular structures indicate the number of adjacent CH groups per
    aromatic ring.}
  \label{FigPAHstructures}
\end{figure}
It is now possible to quantify the relative amounts of the various
types of CH groups on the periphery of the interstellar PAHs which
dominate the emission in this wavelength range and derive the types
and sizes of interstellar PAH structures implied. This is achieved by
analysing the peak positions and integrated band strengths listed in
Table~\ref{TabFeatureProperties} with the laboratory data summarised
in Table~\ref{TabLabBands}. In the following we assume that the 11.2
$\mu$m emission band arises from the the solo vibrations, the 12.0
$\mu$m band from duo modes, the 12.7 $\mu$m from trios and the 13.5
$\mu$m from quartets. From the laboratory data we get the intrinsic
strength of solos relative to the duos, trios and quartets as 6.2, 2.6
and 2.1 respectively (see Table~\ref{TabLabBands}). From the
intensities of the individual features listed in
Table~\ref{TabFeatureProperties} for NGC 7027 we see that the observed
solo to duo, solo to trio and solo to quartet intensity ratios are
23.8, 4.0, 14.7 respectively. We conclude that in NGC 7027 the ratio
of the number of solo CH bonds to CH bonds in duo, trio and quartet
groups are 3.8, 1.5 and 7.1. Taking into account the number of CH's
per functional group there are 7.7, 4.6 and 28.5 solo (groups)
relative to the duo, trio and quartet groups. For IRAS 18317 these
values are (see also Table~\ref{TabAstroRatios}) 3.4, 0.8 and 10.2
respectively. These two sources represent the extremes in the observed
intensity ratio in our sample.

Examples of the types of PAHs which simultaneously satisfy these
different structural constraints are shown in
Fig.~\ref{FigPAHstructures}. In constructing these structures one
should keep in mind that solo H's represent long straight edges, while
the corners in the structures give rise to duos or trios. To match the
dominance of the 11.2 $\mu$m feature in NGC 7027 one is naturally
driven towards rather large molecules with at least $\simeq$100$-$200
carbon atoms and long straight edges. This is entirely in keeping with
previous theoretical calculations of the molecular sizes of the PAH
species which account for most of the emission in these features
\citep{1993ApJ...415..397S}. As illustrated in
Fig.~\ref{FigPAHstructures} structure 1 approximates the ratios listed
in Table~\ref{TabAstroRatios} for NGC 7027, these ratios requires a
preponderance of solo hydrogens over duos and trios. The extreme
quartet to solo group ratio observed for NGC 7027 is not well
reproduced even by figure~\ref{FigPAHstructures} structure 1. In order
to reproduce that ratio in one single molecule one requires to go to
even larger molecules. Rather, we surmise the extreme ratio reflects
the presence of molecules without \emph{any} quartet groups. Quartet
groups represent pendant rings on the molecule, which can be taken off
without altering the other ratios strongly. Of course many other PAH
structures that match the observed ratios are possible. However these
structures are all very similar to this and the conclusion is the
same: in NGC 7027 the PAH family is dominated by large compact PAHs.

In contrast, for IRAS 18317 the situation is very different. The
observed ratios force one to include more corners or uneven edges.
Structure 4 in Fig.~\ref{FigPAHstructures} illustrates one way of
achieving this. Of course this effect can also be achieved by going to
smaller compact structures of the type shown in structure 1 or by
breaking up structure 4 in two or more fragments. Structures 2 and 3
shown in Fig.~\ref{FigPAHstructures} are intermediate between these
two extremes and have solo/duo, solo/trio, and solo/quartet ratios
consistent with the relative interstellar band intensities shown in
Fig.~\ref{FigTo127} for the objects which lie between the extremes,
NGC 7027 and the IRAS 18317. Thus we observe a structural evolution
where closed, compact species dominate the emission in some regions,
while open, uneven structures are more important in others. We surmise
that this structural evolution as revealed by the smooth spectral
evolution shown in Fig.~\ref{FigTo127} reflects the variations in
chemical history and excitation environment in these regions.

\section{Discussion}
\label{SecDiscussion}
In this section we will discuss the assignment of 11.2 and 12.7
interstellar bands and observed variations in 11.2/12.7 $\mu$m ratio
and their correlation with other UIR bands in view of their underlying
physical causes (Sect.~\ref{SecDiscussVariations}). In the past these
variations have been attributed to dehydrogenation. We reconsider this
suggestion in Sect.~\ref{SecDiscussDehydro}.

\subsection{The 11.2,12.7 $\mu$m emission features.}
\label{SecDiscussVariations}
Since quantum calculations show that the CH stretch (3.3 $\mu$m band)
is inherently very weak in ionized PAHs \citep{1996JPC...100.2819L},
the 3.3 $\mu$m emission band is likely a measure of the neutral PAH
contribution. Based on both the peak position of the 11.2 $\mu$m band
(Sect.~\ref{SecCompareLabISM}) and its correlation with the 3.3 $\mu$m
band (Sect.~\ref{SecTrends}) we assign this band to solo CH
out$-$of$-$plane bending vibrations of neutral PAHs. This result means
that in evolved stars the OOP spectrum is dominated by contributions
from the \emph{neutral} PAHs. We already assigned the weak bands at
10.6 and 11.0 $\mu$m to cations (Sect.~\ref{SecCompareLabISM}).  Even
though the absorption cross-sections for the OOP modes do not change
upon ionisation, we cannot directly derive the ionisation fraction
from the measured band strength ratios. This is due to the fact that
ionisation does strongly affect the strength of other modes, which in
turn influences the fraction of absorbed energy emitted at any
wavelength \citep{2000Bakes}.

The spectral identification of the 12.7 band is much less clear and it
is not possible at this time to assign this band unambiguously to
either neutral or cationic PAHs. In the existing database there is
only one species with a strong band that matches well in position,
which `happens' to be a cation. Furthermore the strength of the
interstellar 12.7 $\mu$m band correlates with the strength of the 6.2
$\mu$m feature (see Sect.~\ref{SecTrends}). The strength of the modes
between 6 and 9 $\mu$m are greatly enhanced upon ionisation, and thus,
one way to understand this correlation is to assume that the 12.7 is
also predominantly carried by cations.

Thus, seemingly, the 11.2 $\mu$m and 12.7 $\mu$m bands represent a
dichotomy of interstellar PAHs with the former carried mainly by
neutral and the latter by positively charged PAHs. The origin of this
interrelation between charge and spectral characteristics in unclear.
There is no indication in the laboratory experiments for a causal
relation between, for example, charge state and the relative strength
of the solo to trio modes. Considering also the discussion on the
molecular structures implied by the relative fraction of the solos to
duos and trios (Sect.~\ref{SecMolstructure};
Fig.~\ref{FigPAHstructures}) we are forced to conclude that the good
correlation between the 11.2 and 3.3 $\mu$m bands and between the 12.7
and the 6.2 $\mu$m bands reflects \emph{a correlation of molecular
  structure and charge state with environment}.  Indeed when using
PAHs containing some 50 C-atoms \citep{1984A&A...137L...5L} the
correlation between the $I_\mathrm{11.2}$/$I_\mathrm{6.2}$ and
$I_\mathrm{3.3}$/$I_\mathrm{6.2}$ is well reproduced by model
calculations of \citet{2000Bakes} by only varying the degree of
ionisation. Thus, those environments which favour large PAHs and the
11.2 $\mu$m band (structure 1 in Fig.~\ref{FigPAHstructures}) also
favour neutral PAHs.  While in regions where open uneven molecular
structures and the 12.7 $\mu$m band (structure 4 in
Fig.~\ref{FigPAHstructures}) dominate, PAHs are predominantly charged.

This is probably also the origin of the correlation between the
$I_\mathrm{PAH}$/$I_\mathrm{IR}$ ratio and the
$I_\mathrm{12.7}$/$I_\mathrm{11.2}$ (cf.  Fig.~\ref{FigTrend5}). The
$I_\mathrm{PAH}$/$I_\mathrm{IR}$ measures the PAH/dust abundance
ratio. The loose correlation suggests that for the ISM sources the PAH
abundance is lower. We recognise that PNe inject freshly synthesised
PAHs into the ISM where they are mixed and processed by FUV photons
and shocks. This processing will lead to a slow destruction of the
PAHs.

The dominant molecular structure reflects the integrated history of
the PAH family and we note that all sources with a strong 11.2 $\mu$m
band are PNe, which have formed their PAHs within the last some 1000
years. Because open uneven molecular structures are kinetically more
reactive to the addition of carbon atoms than compact structures, the
predominance of the latter in chemically reactive regions where PAHs
have recently formed can be rationalised.
\citep{1989ApJ...341..372F,1992ApJ...401..269C}. In contrast regions
with relatively strong 12.7 $\mu$m bands are all H~{\sc ii} regions
where luminous stars illuminate material which has been processed in
the ISM for some 10$^9$ years. This processing irreversibly leads to a
breaking down of the molecular structure because reformation is
prohibited by the low temperature of the ISM.

This does not directly explain why the 11.2 $\mu$m band correlates
with the neutral PAH indicator while the 12.7 $\mu$m emission feature
correlates with bands attributed to ions. The charge state is rapidly
set by the charge balance, which is dominated by local physical
conditions, or more specifically the ionisation parameter,
$(G_\mathrm{0}\,\sqrt{T_\mathrm{gas}})/n_\mathrm{e}$, where
$G_\mathrm{0}$ is the FUV radiation field, $T_\mathrm{gas}$ is the
gas-temperature and $n_\mathrm{e}$ the electron density.  Thus rather
than history, ionisation reflects the present. Possibly most of the
destruction is occurring presently and is also driven by local
physical conditions.

\subsection{Dehydrogenation}
\label{SecDiscussDehydro}
We have argued that the smooth changes in the band strength ratios in
the region of interest are caused by variations in the edge-structure
of the dominant emitting species. However other effects can also be of
influence on the emitted spectra. Most notably dehydrogenation. There
has been a long debate in the literature on the effect of
dehydrogenation on the spectral characteristics of the 10$-$15 $\mu$m
region. Originally when only the 11.2 UIR band was known, its
dominance had been attributed to extreme ($\sim$90\%) dehydrogenation
of the emitting aromatic species leaving only solo hydrogens
\citep{1981MNRAS.196..269D}. However, this question was revisited when
IRAS revealed the presence of duos and trios in the interstellar PAH
family \citep{1985ApJ...299L..93C}. Theoretical studies of the
dehydrogenation of interstellar PAHs have shown that for PAHs larger
than about 25 C-atoms hydrogenation through reactions with abundant
atomic H is more important than H loss through unimolecular
dissociation \citep{1987paha.proc..273T,ATB89, 1994ApJ...420..307J,
  1996A&A...305..616A, 1999ApJ...512..500J}.  Hence, with a typical
PAH size of 50 C-atoms dehydrogenation should have no effect on the
UIR spectrum.

Observationally, our analysis also argues against dehydrogenation.
First, we observe a constant ratio of the 3.3 $\mu$m band (all CH
oscillation) to the 11.2 $\mu$m band (only solo CH oscillation).
However, we would expect a non-linear behaviour since, when
dehydrogenation commences the number of solo H increases as duos and
trios are converted to solo's and only at high dehydrogenation does
the relation between the 3.3 and the 11.2 $\mu$m bands become linear
\citep{1993ApJ...415..397S}. Secondly, if the variation in
$I_\mathrm{12.7}$/$I_\mathrm{11.2}$ reflects dehydrogenation than we
would expect that decreasing H coverage (ie. decreasing
$I_\mathrm{12.7}$/$I_\mathrm{11.2}$) would correlate with increasing
CC/CH mode emission (ie.  $I_\mathrm{6.2}$/$I_\mathrm{11.2}$). The
opposite is actually observed (cf.  Fig.~\ref{FigTrend2}).  We
conclude therefore that dehydrogenation has little influence on the
observed interstellar UIR spectrum.

\section{Summary}
\label{SecSummary}
We have presented new 10$-$15 $\mu$m spectra of evolved stars, H~{\sc
  ii} regions, RNe and YSOs. We observe very rich UIR spectra with
strong bands at 11.2 and 12.7 $\mu$m and weaker bands at 10.6, 11.0,
12.0, 13.5 and 14.2 $\mu$m. These spectra show large variations in the
band strength ratios between sources, especially in the 11.2 and 12.7
$\mu$m feature ratio. Evolved stars have a dominant 11.2 $\mu$m
feature while in H~{\sc ii} regions the 12.7 and 11.2 are typically
equally strong. We find that the 11.2 $\mu$m band correlates with the
CH stretch band at 3.3 $\mu$m and that the 12.7 $\mu$m band correlates
with the CC stretch band at 6.2 $\mu$m.

We have summarised new laboratory spectroscopy results for the CH
out$-$of$-$plane bending vibrations on isolated neutral and cationic
PAHs.  Different number of adjacent CH bonds give rise to vibrations
in distinctly different wavelength regions. The modes are therefore
good diagnostics of the molecular structure of the emitting species.
Upon ionisation the solo CH vibrations are shifted to shorter
wavelength compared to the solo modes in neutral. The cross-sections
per mode are not strongly modified upon ionisation. We attribute the
weak bands at 10.6 and 11.0 $\mu$m to solo modes in positively charged
PAHs, the strong 11.2 $\mu$m band the solo modes in neutral. The weak
12.0 $\mu$m band we assign to the duo modes, the 12.7 $\mu$m to trio
modes and the 13.5 $\mu$m feature to quartet vibrations.

From the average cross-sections per mode we have constrained the
relative numbers of solo, duo, trio and quartet CH groups in different
sources for the PAH species that effectively emit in this wavelength
region. The spectra of PNe with a dominant 11.2 $\mu$m feature arises
from large ($\sim$ 100$-$150 C-atom) compact PAHs with long straight
edges. In contrast the H~{\sc ii} region spectra are due to smaller or
more irregular PAHs.

We propose a scenario in which large compact PAHs are formed in the
winds around evolved stars. These PAHs are consequently degraded in
the ISM. From the correlations between charge indicators, which are
set by the local physical conditions, and the 11.2/12.7 $\mu$m band
strength ratio, which is determined by the molecular structure, we
conclude that much of this degradation happens on a short timescale in
the emission objects itself.

\begin{acknowledgements}
  The authors wish to thank the referee dr. L. Verstraete whose
  comments have helped to improve the paper. SH acknowledges the
  support from an NWO program, grant 616-78-333.  EP acknowledges the
  support from an NWO program, grant 783-70-000.  CVK is a Research
  Assistant of the Fund for Scientific Research. DMH and LJA
  gratefully acknowledge support under NASA's IR Laboratory
  Astrophysics (344-02-06-01) and Long Term Space Astrophysics
  programs (399-20-01). IA$^3$ is a joint development of the SWS
  consortium. Contributing institutes are SRON, MPE, KUL and the ESA
  Astrophysics Division. This work was supported by the Dutch ISO Data
  Analysis Center(DIDAC). The DIDAC is sponsored by SRON, ECAB, ASTRON
  and the universities of Amsterdam, Groningen, Leiden and Leuven.
\end{acknowledgements}

\bibliographystyle{apj}
\bibliography{articles}

\begin{thebibliography}{47}
\expandafter\ifx\csname natexlab\endcsname\relax\def\natexlab#1{#1}\fi

\bibitem[{{Allain} {et~al.}(1996){Allain}, {Leach}, \&
  {Sedlmayr}}]{1996A&A...305..616A}
{Allain}, T., {Leach}, S., \& {Sedlmayr}, E. 1996, \aap, 305, 616

\bibitem[{{Allamandola} {et~al.}(1999){Allamandola}, {Hudgins}, \&
  {Sandford}}]{1999ApJ...511L.115A}
{Allamandola}, L.~J., {Hudgins}, D.~M., \& {Sandford}, S.~A. 1999, \apjl, 511,
  L115

\bibitem[{{Allamandola} {et~al.}(1985){Allamandola}, {Tielens}, \&
  {Barker}}]{ATB85}
{Allamandola}, L.~J., {Tielens}, A. G. G.~M., \& {Barker}, J.~R. 1985, \apjl,
  290, L25

\bibitem[{{Allamandola} {et~al.}(1989){Allamandola}, {Tielens}, \&
  {Barker}}]{ATB89}
{Allamandola}, L.~J., {Tielens}, G. G.~M., \& {Barker}, J.~R. 1989, \apjs, 71,
  733

\bibitem[{{Bakes} {et~al.}(2000)}]{2000Bakes}
{Bakes}, E. {et~al.} 2000, \apj, in preparation

\bibitem[{Bellamy(1958)}]{1958Bellamy}
Bellamy, L. 1958, The infra-red spectra of complex molecules, 2nd ed. (Wiley:
  New York)

\bibitem[{{Boulanger} {et~al.}(1998){Boulanger}, {Boisssel}, {Cesarsky}, \&
  {Ryter}}]{1998A&A...339..194B}
{Boulanger}, F., {Boisssel}, P., {Cesarsky}, D., \& {Ryter}, C. 1998, \aap,
  339, 194

\bibitem[{{Brenner} \& {Barker}(1992)}]{1992ApJ...388L..39B}
{Brenner}, J. \& {Barker}, J.~R. 1992, \apjl, 388, L39

\bibitem[{{Cesarsky} {et~al.}(1996){Cesarsky}, {Abergel}, {Agnese}, {Altieri},
  {Augueres}, {Aussel}, {Biviano}, {Blommaert}, {Bonnal}, {Bortoletto},
  {Boulade}, {Boulanger}, {Cazes}, {Cesarsky}, {Chedin}, {Claret}, {Combes},
  {Cretolle}, {Davies}, {Desert}, {Elbaz}, {Engelmann}, {Epstein},
  {Franceschini}, {Gallais}, {Gastaud}, {Gorisse}, {Guest}, {Hawarden},
  {Imbault}, {Kleczewski}, {Lacombe}, {Landriu}, {Lapegue}, {Lena}, {Longair},
  {Mandolesi}, {Metcalfe}, {Mosquet}, {Nordh}, {Okumura}, {Ott}, {Perault},
  {Perrier}, {Persi}, {Puget}, {Purkins}, {Rio}, {Robert}, {Rouan}, {Roy},
  {Saint-Pe}, {Sam Lone}, {Sargent}, {Sauvage}, {Sibille}, {Siebenmorgen},
  {Sirou}, {Soufflot}, {Starck}, {Tiphene}, {Tran}, {Ventura}, {Vigroux},
  {Vivares}, \& {Wade}}]{1996A&A...315L..32C}
{Cesarsky}, C.~J., {Abergel}, A., {Agnese}, P., {et~al.} 1996, \aap, 315, L32

\bibitem[{{Chan} {et~al.}(2000){Chan}, {Roellig}, {Onaka}, {Mizutani},
  {Okumura}, {Yamamura}, {Tanab{\'e}}, {Shibai}, {Nakagawa}, \&
  {Okuda}}]{2000ibp..confE..14C}
{Chan}, K., {Roellig}, T.~L., {Onaka}, T., {et~al.} 2000, in ISO beyond the
  peaks: The 2nd ISO workshop on analytical spectroscopy, held 2-4 February
  2000, at VILSPA., E14

\bibitem[{{Cherchneff} {et~al.}(1992){Cherchneff}, {Barker}, \&
  {Tielens}}]{1992ApJ...401..269C}
{Cherchneff}, I., {Barker}, J.~R., \& {Tielens}, A. G. G.~M. 1992, \apj, 401,
  269

\bibitem[{{Cohen} {et~al.}(1986){Cohen}, {Allamandola}, {Tielens}, {Bregman},
  {Simpson}, {Witteborn}, {Wooden}, \& {Rank}}]{1986ApJ...302..737C}
{Cohen}, M., {Allamandola}, L., {Tielens}, A. G. G.~M., {et~al.} 1986, \apj,
  302, 737

\bibitem[{{Cohen} {et~al.}(1985){Cohen}, {Tielens}, \&
  {Allamandola}}]{1985ApJ...299L..93C}
{Cohen}, M., {Tielens}, A. G. G.~M., \& {Allamandola}, L.~J. 1985, \apjl, 299,
  L93

\bibitem[{{Cohen} {et~al.}(1989){Cohen}, {Tielens}, {Bregman}, {Witteborn},
  {Rank}, {Allamandola}, {Wooden}, \& {Jourdain De
  Muizon}}]{1989ApJ...341..246C}
{Cohen}, M., {Tielens}, A. G. G.~M., {Bregman}, J., {et~al.} 1989, \apj, 341,
  246

\bibitem[{{Colangeli} {et~al.}(1992){Colangeli}, {Mennella}, \&
  {Bussoletti}}]{1992ApJ...385..577C}
{Colangeli}, L., {Mennella}, V., \& {Bussoletti}, E. 1992, \apj, 385, 577

\bibitem[{{Cook} \& {Saykally}(1998)}]{1998ApJ...493..793C}
{Cook}, D.~J. \& {Saykally}, R.~J. 1998, \apj, 493, 793

\bibitem[{{Cox} \& {Kessler}(1999)}]{1999ESASP.427.....C}
{Cox}, P. \& {Kessler}, M. 1999, ESA SP-427: The Universe as Seen by ISO, 427

\bibitem[{{de Graauw} {et~al.}(1996){de Graauw}, {Haser}, {Beintema},
  {et~al.}}]{1996A&A...315L..49D}
{de Graauw}, T., {Haser}, L.~N., {Beintema}, D.~A., {et~al.} 1996, \aap, 315,
  L49

\bibitem[{{Duley} \& {Williams}(1981)}]{1981MNRAS.196..269D}
{Duley}, W.~W. \& {Williams}, D.~A. 1981, \mnras, 196, 269

\bibitem[{{Flickinger} {et~al.}(1991){Flickinger}, {Wdowiak}, \&
  {Gomez}}]{1991ApJ...380L..43F}
{Flickinger}, G.~C., {Wdowiak}, T.~J., \& {Gomez}, P.~L. 1991, \apjl, 380, L43

\bibitem[{{Frenklach} \& {Feigelson}(1989)}]{1989ApJ...341..372F}
{Frenklach}, M. \& {Feigelson}, E.~D. 1989, \apj, 341, 372

\bibitem[{{Gesicki} \& {Zijlstra}(2000)}]{2000A&A...358.1058G}
{Gesicki}, K. \& {Zijlstra}, A.~A. 2000, \aap, 358, 1058

\bibitem[{{Habing}(1968)}]{1968BAN....19..421H}
{Habing}, H.~J. 1968, \bain, 19, 421

\bibitem[{{Hudgins} \& {Allamandola}(1999)}]{1999ApJ...516L..41H}
{Hudgins}, D.~M. \& {Allamandola}, L.~J. 1999, \apjl, 516, L41

\bibitem[{{Hudgins} {et~al.}(2000{\natexlab{a}}){Hudgins}, {Bauschlicher}, \&
  {Allamandola}}]{2000HudginsB}
{Hudgins}, D.~M., {Bauschlicher}, C.~W., J., \& {Allamandola}, L.~J.
  2000{\natexlab{a}}, Jour. Phys. Chem. A, 104, 3655

\bibitem[{{Hudgins} {et~al.}(2000{\natexlab{b}}){Hudgins}, {Bauschlicher}, \&
  {Allamandola}}]{2000Hudgins}
---. 2000{\natexlab{b}}, \apj, in preparation

\bibitem[{{Joblin} {et~al.}(1995){Joblin}, {Boissel}, {Leger}, {D'Hendecourt},
  \& {Defourneau}}]{1995A&A...299..835J}
{Joblin}, C., {Boissel}, P., {Leger}, A., {D'Hendecourt}, L., \& {Defourneau},
  D. 1995, \aap, 299, 835

\bibitem[{{Jochims} {et~al.}(1999){Jochims}, {Baumg\"artel}, \&
  {Leach}}]{1999ApJ...512..500J}
{Jochims}, H.~W., {Baumg\"artel}, H., \& {Leach}, S. 1999, \apj, 512, 500

\bibitem[{{Jochims} {et~al.}(1994){Jochims}, {Ruhl}, {Baumgartel}, {Tobita}, \&
  {Leach}}]{1994ApJ...420..307J}
{Jochims}, H.~W., {Ruhl}, E., {Baumgartel}, H., {Tobita}, S., \& {Leach}, S.
  1994, \apj, 420, 307

\bibitem[{{Kessler} {et~al.}(1996){Kessler}, {Steinz}, {Anderegg},
  {et~al.}}]{1996A&A...315L..27K}
{Kessler}, M.~F., {Steinz}, J.~A., {Anderegg}, M.~E., {et~al.} 1996, \aap, 315,
  L27

\bibitem[{{Langhoff}(1996)}]{1996JPC...100.2819L}
{Langhoff}, S.~R. 1996, Jour. Phys. Chem., 100, 2819

\bibitem[{{Latter} {et~al.}(2000){Latter}, {Dayal}, {Bieging}, {Meakin},
  {Hora}, {Kelly}, \& {Tielens}}]{2000ApJ...539..783L}
{Latter}, W.~B., {Dayal}, A., {Bieging}, J.~H., {et~al.} 2000, \apj, 539, 783

\bibitem[{{Leger} {et~al.}(1989){Leger}, {D'Hendecourt}, \&
  {Defourneau}}]{1989A&A...216..148L}
{Leger}, A., {D'Hendecourt}, L., \& {Defourneau}, D. 1989, \aap, 216, 148

\bibitem[{{Leger} \& {Puget}(1984)}]{1984A&A...137L...5L}
{Leger}, A. \& {Puget}, J.~L. 1984, \aap, 137, L5

\bibitem[{{Moutou} {et~al.}(2000){Moutou}, {Verstraete}, {Leger}, {Sellgren},
  \& {Schmidt}}]{2000A&A...354L..17M}
{Moutou}, C., {Verstraete}, L., {Leger}, A., {Sellgren}, K., \& {Schmidt}, W.
  2000, \aap, 354, L17

\bibitem[{{Piest} {et~al.}(1999){Piest}, {von Helden}, \&
  {Meijer}}]{1999ApJ...520L..75P}
{Piest}, H., {von Helden}, G., \& {Meijer}, G. 1999, \apjl, 520, L75

\bibitem[{{Roche} {et~al.}(1989){Roche}, {Aitken}, \&
  {Smith}}]{1989MNRAS.236..485R}
{Roche}, P.~F., {Aitken}, D.~K., \& {Smith}, C.~H. 1989, \mnras, 236, 485

\bibitem[{{Schutte} {et~al.}(1993){Schutte}, {Tielens}, \&
  {Allamandola}}]{1993ApJ...415..397S}
{Schutte}, W.~A., {Tielens}, A. G. G.~M., \& {Allamandola}, L.~J. 1993, \apj,
  415, 397

\bibitem[{{Szczepanski} {et~al.}(1995){Szczepanski}, {Drawdy}, {Wehlburg}, \&
  {Vala}}]{1995CPL...245..539S}
{Szczepanski}, J., {Drawdy}, J., {Wehlburg}, C., \& {Vala}, M. 1995, Chem.
  Phys. Lett., 245, 539

\bibitem[{{Tielens} {et~al.}(1987){Tielens}, {Allamandola}, {Barker}, \&
  {Cohen}}]{1987paha.proc..273T}
{Tielens}, A. G. G.~M., {Allamandola}, L.~J., {Barker}, J.~R., \& {Cohen}, M.
  1987, in Polycyclic Aromatic Hydrocarbons and Astrophysics, 273--285

\bibitem[{{Uchida} {et~al.}(2000){Uchida}, {Sellgren}, {Werner}, \&
  {Houdashelt}}]{2000ApJ...530..817U}
{Uchida}, K.~I., {Sellgren}, K., {Werner}, M.~W., \& {Houdashelt}, M.~L. 2000,
  \apj, 530, 817

\bibitem[{{Van Kerckhoven} {et~al.}(2000){Van Kerckhoven}, {Hony}, {Peeters},
  {Tielens}, {Allamandola}, {Hudgins}, {Cox}, {Roelfsema}, {Voors}, {Waelkens},
  {Waters}, \& {Wesselius}}]{2000A&A...357.1013V}
{Van Kerckhoven}, C., {Hony}, S., {Peeters}, E., {et~al.} 2000, \aap, 357, 1013

\bibitem[{{Verstraete} {et~al.}(1996){Verstraete}, {Puget}, {Falgarone},
  {Drapatz}, {Wright}, \& {Timmermann}}]{1996A&A...315L.337V}
{Verstraete}, L., {Puget}, J.~L., {Falgarone}, E., {et~al.} 1996, \aap, 315,
  L337

\bibitem[{{Verstraete} {et~al.}(2001)}]{2001Verstraete}
{Verstraete}, L. {et~al.} 2001, \aap, in press

\bibitem[{{Waters} {et~al.}(1998){Waters}, {Cami}, {De Jong}, {Molster}, {Van
  Loon}, {Bouwman}, {De Koter}, {Waelkens}, {Van Winckel}, \&
  {Morris}}]{1998Natur.391..868W}
{Waters}, L. B. F.~M., {Cami}, J., {De Jong}, T., {et~al.} 1998, \nat, 391, 868

\bibitem[{{Witteborn} {et~al.}(1989){Witteborn}, {Sandford}, {Bregman},
  {Allamandola}, {Cohen}, {Wooden}, \& {Graps}}]{1989ApJ...341..270W}
{Witteborn}, F.~C., {Sandford}, S.~A., {Bregman}, J.~D., {et~al.} 1989, \apj,
  341, 270

\bibitem[{{Wolfire} {et~al.}(1989){Wolfire}, {Hollenbach}, \&
  {Tielens}}]{1989ApJ...344..770W}
{Wolfire}, M.~G., {Hollenbach}, D., \& {Tielens}, A. G. G.~M. 1989, \apj, 344,
  770

\end{thebibliography}

\end{document}